\documentclass[twocolumn,showpacs,aps,pra]{revtex4-1}

\newcommand{\dd}{\mathrm d}
\newcommand{\ee}{\mathrm e}
\newcommand{\ii}{\mathrm i}

\newcommand{\calE}{\mathcal E}
\newcommand{\calF}{\mathcal F}

\newcommand{\calP}{\mathcal P}
\newcommand{\calL}{\mathcal L}

\newcommand{\calS}{\mathcal S}
\newcommand{\calT}{\mathcal T}
\newcommand{\calZ}{\mathcal Z}
\newcommand{\Ci}{{\mathrm{Ci}}}
\newcommand{\Si}{{\mathrm{Si}}}
\newcommand{\sgn}{\varepsilon}

\usepackage{epsf}
\usepackage{amsmath}
\usepackage{amsfonts}
\usepackage{amssymb}
\usepackage{bm}
\usepackage{bbm}
\usepackage{nicefrac}
\usepackage{color}
\usepackage{pifont}
\usepackage{graphicx}
\usepackage{enumerate}
\usepackage{dcolumn}
\usepackage{maybemath}
\usepackage{xfrac}

\definecolor{kellygreen}{rgb}{0.3, 0.73, 0.09}
\definecolor{duelferred}{rgb}{0.7, 0.2, 0.1}
\definecolor{garrosgreen}{rgb}{0.1, 0.4, 0.1}
\definecolor{cambridgeblue}{rgb}{0.1, 0.3, 1.0}

\begin{document}

\newcommand{\addrMST}{
Department of Physics, Missouri University of Science and Technology,
Rolla, Missouri 65409, USA}

\title{Long--Range Atom--Wall Interactions and Mixing Terms: Metastable Hydrogen}

\author{U.~D.~Jentschura}

\affiliation{\addrMST}

\begin{abstract}
We investigate the interaction of metastable $2S$ hydrogen atoms with a
perfectly conducting wall, including parity-breaking $S$--$P$ mixing terms
(with full account of retardation). The neighboring $2P_{1/2}$ and $2P_{3/2}$
levels are found to have a profound effect on the transition from the
short-range, nonrelativistic regime, to the retarded form of the
Casimir--Polder interaction.  The corresponding $P$ state admixtures to the
metastable $2S$ state are calculated.  We find the long-range asymptotics of
the retarded Casimir--Polder potentials and mixing amplitudes, for general
excited states, including a fully quantum electrodynamic treatment of the
dipole-quadrupole mixing term.  The decay width of the metastable $2S$ state is
roughly doubled even at a comparatively large distance of $918\,$atom units
(Bohr radii) from the perfect conductor.  The magnitude of the calculated
effects is compared to the unexplained Sokolov effect.
\end{abstract}


\maketitle

{\em Introduction.---}The 
investigation of atom-wall interactions for atoms in contact with
conducting materials has a long history. Starting from the works of 
Lennard--Jones~\cite{LJ1932}, Bardeen~\cite{Ba1940}, Casimir and
Polder~\cite{CaPo1948}, and Lifshitz~\cite{Li1955}, research on related matters
has found continuously growing interest over the last
decades~\cite{WySi1984,WySi1985,BaDo1987,BaDo1990}.  In the non-retarded regime
(close range), the interaction energy scales as $1/\calZ^3$ with the atom-wall
distance $\calZ$, while for atom-wall distances large in comparison to a
typical atomic wavelength, the interaction energy scales as $1/\calZ^4$ (see
Chap.~8 of Ref.~\cite{Mi1994}).  The leading term is given by virtual dipole
transitions, while multipole corrections have recently been analyzed in
Ref.~\cite{LaDKJe2010pra}.  The symmetry breaking induced by the wall leads to
dipole-quadrupole mixing terms, which lead to admixtures to metastable
levels~\cite{BoEtAl2001,KaEtAl2006epl}.  While this effect has been analyzed in
the non-retarded van-der-Waals regime~\cite{BoEtAl2001,KaEtAl2006epl}, a fully
quantum electrodynamic calculation of this effect would be of obvious interest.

This fact is emphasized by the curious 
observation of a long-range, and conceivably super-long-range
(micrometer-scale) interaction of metastable hydrogen $2S$ 
atoms with a conducting surface (the so-called 
Sokolov effect, see Refs.~\cite{SoYaPa1994,KaKaKuPoSo1996,SoYaPaPc2002,SoEtAl2005}).
It is not far-fetched to suspect that 
this effect could be due to a quantum electrodynamically
induced tail of the dipole-quadrupole mixing term
in the atom-wall interaction. Namely,
for the hydrogen $2S$ atom, 
the neighboring $2P_{1/2}$ and $2P_{3/2}$ 
levels are removed only by the Lamb shift and fine-structure,
respectively, while it is known that 
virtual states of lower energy can induce 
long-range tails in atom-wall interactions, as well as in the Lamb shift between plates
(see Refs.~\cite{Ba1970,BaBa1972a,BaBa1972b,Ba1974,Ba1979,%
Fo1979,Ba1987a,Ba1987b,Ba1988,Ba1997}).
The large admixtures typically induced in atomic 
systems when a metastable level couples
to nearly degenerate states of opposite 
parity suggest that a closer investigation of the hydrogen 
system is warranted.
Atomic units with $\hbar = 4\pi\epsilon_0 = 1$ and 
$c = 1/\alpha$ are used throughout this 
Rapid Communication, where $\alpha$ is the fine-structure constant.
The electron charge is explicitly denoted as $e$ unless stated 
otherwise.

%
%
{\em Retardation of the atom-wall interaction.---}The 
quantum electrodynamic (QED) length-gauge interaction
\begin{equation}
\label{HI}
H_I = -e \, \vec r \cdot \vec E - 
\frac{e}{2} r^i \, r^j \, \partial E^i/\partial r^j + \dots \,,
\end{equation}
follows naturally from the 
formalism of long-wavelength QED
interaction Hamiltonian~\cite{PoZi1959,Pa2004}
($\vec r$ denotes the electron coordinate).
In contrast to the vector potential,
the electric field strength (operator) is 
gauge-invariant (this point has given rise 
to some discussion, see Ref.~\cite{Ko1978prl})
and reads as 
[cf.~Eq.~(2.3) of Ref.~\cite{Ba1974}]
\begin{align}
\vec E(\vec r) = & \;
\int\limits_0^\infty \dd L 
\int\limits_{\mathbbm{R}^2} \frac{ \dd^2 k_\parallel }{\pi} \sqrt{\omega}
\left\{ a_1(\vec k, L) 
( \hat{k}_\parallel \times {\hat e}_z )  \sin(L z)
\right.
\nonumber\\[2.0ex]
& \; \left. + a_2(\vec k, L) \, \left[ 
{\hat k}_\parallel \, \frac{\ii L}{\omega} \sin(L z) -
{\hat e}_z \, \frac{k_\parallel}{\omega} \cos(L z) \right] \right\}
\nonumber\\[2.0ex]
& \; \times \ee^{\ii \vec k_\parallel \cdot \vec r_\parallel}
+ {\rm h.c.}\,,
\end{align}
where $\vec r = \vec r_\parallel + z\, \hat e_z$ with 
$\vec r_\parallel = x \, \hat e_x + y \, \hat e_y$,
while $\vec k_\parallel = k_x \, \hat e_x + k_y \, \hat e_y$,
also $\vec k_\perp = k_z \, \hat e_z$,
and $L \equiv | \vec k_\perp |$.
The commutator relation is
$[ a_s(\vec k_\parallel, L), \,
a^\dagger_{s'}(\vec k_\parallel, L) ] =
\delta_{ss'} \,
\delta^{(2)}(\vec k_\parallel -  \vec k'_\parallel) \,
\delta(L - L')$ for the annihilation and creation operators $a_s$ and $a^\dagger_s$.
In order to evaluate the interaction Hamiltonian \eqref{HI},
one shifts $z \to \calZ + z$ where $\calZ$ is the coordinate of the 
atom's center (nucleus).
The proton is at $(0,0,
\calZ)$, while the atomic electron coordinates are $(x,y,\calZ+z)$. 
The surface of the perfect conductor is in the $xy$ plane, i.e., in the plane
described by the points~$(x, y, 0)$. 
The unperturbed Hamiltonian $H_0$
is the sum of the free radiation field and the unperturbed atom
[see Eq.~(2.1) of Ref.~\cite{Ba1974} and Eq.~(3.2) of Ref.~\cite{JeKe2004aop}].
For a reference ground state $| n\rangle$, 
second-order perturbation theory leads to a known 
result given in Eq.~(8.41) of Ref.~\cite{Mi1994}
or Eq.~(27) of Ref.~\cite{LaDKJe2010pra}, which
involves the symmetric sum with imaginary frequency in the argument of the 
dynamics polarizability $\Pi(\pm \ii \omega)$.
The Wick rotation of the virtual photon integration contour,
leads to the symmetrization $\ii
\omega \leftrightarrow  -\ii \omega$ but cannot be done
for excited reference states.
We use second-order perturbation theory 
to evaluate $\Delta E = \langle n | (-e \, \vec r \cdot \vec E) \,
[1/(E_n - H_0')] \,  (-e \, \vec r \cdot \vec E) | n \rangle$
and obtain [cf.~the discussion
following Eq.~(2.12) of Ref.~\cite{Ba1974}],
\begin{widetext}
\begin{align}
\Delta E \doteq 
\frac{e^2}{2 \pi} \, 
{\mbox{(P.V.)}} \, \sum_q 
\int\limits_0^\infty \dd L \,
\int\limits_{L}^\infty \dd \omega \,
\cos(2 L \, \calZ) \,
\frac{ L^2 \,
\left( \left| \langle n | {\vec r}_\parallel | q \rangle \right|^2 +
2 \, \left| \langle n | z | q \rangle \right|^2 \right) 
+ \omega^2 \left( \left| \langle n | {\vec r}_\parallel | q \rangle \right|^2 -
2 \, \left| \langle n | z | q \rangle \right|^2 \right) }%
{\calE_q + \omega - \ii \, \delta} \,,
\end{align}
where the identity $\int_{\mathbbm{R}^3} \dd^3 k = 
2 \int_0^{2 \pi} \dd \varphi 
\int_0^\infty \dd L
\int_0^\infty \dd \omega \, \omega$ \,,
with $\omega = \sqrt{\vec k_\parallel^2 + k_z^2}$,
and $L = |k_z|$ has been  
used in order to transform the integration measure.
The virtual states are denoted as $| q \rangle$,
and their energy difference to the reference state is 
denoted as $\calE_q \equiv E_q - E_n$.
In contrast to the velocity gauge~\cite{Ba1974}, there
is no seagull term to consider, and it is not necessary 
to add the electrostatic interaction with the mirror
charges by hand~\cite{Le1981qed}.
It is an in principle well known (see the remarks following
Eq.~(A.22) in Appendix~A of Ref.~\cite{Sa1967Adv}), 
but sometimes forgotten wisdom that the
Coulomb interaction does not need to be quantized in the
velocity gauge~\cite{Le1981qed}.
The integration with respect to $\omega$ leads to logarithmic 
terms [see the Appendix of Ref.~\cite{Ba1974}].
After the subtraction of $\calZ$-independent terms 
(the subtraction is denoted by the $\doteq$ sign), one obtains
\begin{subequations}
\label{I1I2}
\begin{align}
I_1(\chi) \doteq & \; \int_0^\infty \dd L \, \cos(2 L \calZ) \, \ln( | \calE_q + L | ) = 
\calE_q \, \left( \frac{\pi \, [1 - \sgn(\calE_q)]}{2 \chi} 
- \frac{T(\chi)}{\chi} - \pi \Theta(-\calE_q) \, \frac{2 \sin^2(\tfrac{\chi}{2})}{\chi} \right) \,,
\\[2.0ex]
I_2(\chi) \doteq & \;
-\frac{\partial^2 I_1}{\partial \chi^2} =
\calE_q \, \left( \frac{\pi \, [\sgn(\calE_q)-1] + \chi}{\chi^3} 
+ \frac{2 - \chi^2}{\chi^3} \, T(\chi) 
- \frac{2}{\chi^2} \, U(\chi)
+ \pi  \Theta(-\calE_q) \,
\frac{\partial^2}{\partial \chi^2} \frac{2 \sin^2(\tfrac{\chi}{2})}{\chi} \right) \,,
\\[2.0ex]
T(\chi) =& \; \sin(\chi) \, \Ci(\chi) - \cos(\chi) \, \Si(\chi) + \frac{\pi}{2} \cos(\chi) \,,
\qquad
\chi = 2 | \calE_q| \, \calZ \,,
\qquad
\varepsilon(\calE_q) = 
\Theta(\calE_q) - \Theta(-\calE_q) \,.
\end{align}
\end{subequations}
Here, $\Ci(\chi) = - \int_\chi^\infty \dd t \, \frac{\cos(t)}{t}$
and $\Si(\chi) = \int_0^\chi \dd t \, \frac{\sin(t)}{t}$,
and $U(\chi) = \frac{\partial}{\partial \chi} T(\chi)$,
while $T(\chi) = \chi^{-1} -\frac{\partial}{\partial \chi} U(\chi)$.
We confirm the result given in Eq.~(2.18) of Ref.~\cite{Ba1974}
and represent 
the ``distance-dependent Lamb shift'' as 
\begin{align}
\label{RESH}
\Delta E \doteq & \frac{e^2}{2 \pi} \sum_q \calE_q^3 \,
\biggl\{ \left( \left| \langle n | {\vec r}_\parallel | q \rangle \right|^2 -
2 \, \left| \langle n | z | q \rangle \right|^2 \right) \,
\left[ \frac{\pi \, [\sgn(\calE_q) - 1]}{2 \chi} - \frac{1}{\chi^2} + 
\frac{T(\chi)}{\chi} 
+ \pi \Theta(-\calE_q) \, \frac{1 - \cos(\chi)}{\chi} \right]
\\[2.0ex]
& \; 
- \left( \left| \langle n | {\vec r}_\parallel | q \rangle \right|^2 +
2 \left| \langle n | z | q \rangle \right|^2 \right) \,
\left[ \frac{\pi \, [\sgn(\calE_q)-1] + \chi}{\chi^3} 
+ \frac{2 - \chi^2}{\chi^3} \, T(\chi)
- \frac{2}{\chi^2} \, U(\chi)
+ \pi \Theta(-\calE_q) 
\frac{\partial^2}{\partial \chi^2} 
\frac{1- \cos(\chi)}{\chi} \right] \biggr\}  \,.
\nonumber
\end{align}
We should perhaps clarify that 
the $\calZ$-independent contribution to the 
Lamb shift (the ordinary ``free-space Lamb shift'') is
absorbed in the subtraction procedure denoted here by
the ``$\doteq$'' sign in Eqs.~\eqref{I1I2},~\eqref{RESH},~\eqref{J1J2} and~\eqref{RESM}.
The $\calZ$-dependent position of the energy level is obtained after adding the
``free-space Lamb shift''~$\calL$ and ``free-space fine structure''~$\calF$
given in Eq.~\eqref{LF} to the $\calZ$-dependent energy shifts
given in Eqs.~\eqref{RESH} and~\eqref{RESM}.
In the nonretardation limit, the
$\calZ$-dependent results given in Eqs.~\eqref{RESH} and~\eqref{RESM}
are replaced by the respective terms of the nonretarded potential~\eqref{VVV}.
This (somewhat subtle) point is not fully discussed in previous
works on the subject~\cite{Ba1970,BaBa1972a,BaBa1972b,Ba1974,Ba1979}
and therefore should be mentioned for absolute clarity.

The term $-\chi^{-2}$ in the coefficient multiplying 
$\left| \langle n | {\vec r}_\parallel | q \rangle \right|^2 -
2 \, \left| \langle n | \vec z | q \rangle \right|^2$
vanishes after summing over the entire spectrum of virtual states;
it is obtained naturally in the length gauge 
and otherwise cancels a term in the expansion of the 
energy shift for large $\chi$ (even before the application of the sum rule,
which is crucial in velocity gauge~\cite{Ba1974}).
The off-diagonal mixing term leads to 
to the matrix element 
$\Delta M = \langle m | (-e\, \vec r \cdot \vec E) \, [1/(E_n - H_0)'] 
(- \frac{e}{2} r^i \, r^j \, (\partial E^i/\partial r^j) | n \rangle + 
\langle m | {\rm h.c.} | n \rangle$,
\begin{subequations}
\begin{align}
\Delta M =& \;
\frac{e^2}{4 \pi}
{\mbox{(P.V.)}} \, \sum_q 
\int\limits_0^\infty \dd L \,
\int\limits_{L}^\infty \dd \omega \,
\frac{L \, \sin(2 \, L \, \calZ)}{\calE_q + \omega - \ii \delta} \,
\left( L^2 \, \langle n | \calT_2 | m \rangle- 
\omega^2 \, \langle n | \calT_1 | m \rangle \right) \,,
\\[2.0ex]
\langle m | \calT_1 | n \rangle =& \; \langle m | z | q \rangle \,
\langle q | {\vec r}^{\,2}_\parallel - 2\, z^2 | n \rangle 
+ \langle m | {\rm h.c.} | n \rangle  \,,
\\[1.133ex]
\langle m | \calT_2 | n \rangle =& \;
\langle m | z  | q \rangle \,
\langle q | {\vec r}^{\,2}_\parallel - 2\, z^2 | n \rangle -
2 \, \langle m | {\vec r}_\parallel | q \rangle \cdot
\langle q | {\vec r}_\parallel \, z | n \rangle 
+ \langle m | {\rm h.c.} | n \rangle \,.
\end{align}
\end{subequations}
After the subtraction of $\calZ$-independent terms,
the following two results for 
$J_1(\chi) = \int_0^\infty \dd L L \sin(2 L \calZ) \ln( | \calE_q + L | )$ and 
$J_2(\chi) = -\partial^2 J_1(\chi)/\partial\chi^2$
supplement the analytic integrals given in Eq.~\eqref{I1I2},
\begin{subequations}
\label{J1J2}
\begin{align}
J_1(\chi) \doteq & \; 
\calE_q^2 \left( 
\sgn(\calE_q) 
\left( \frac{\pi}{2 \chi^2} - \frac{T(\chi)}{\chi^2} + \frac{U(\chi)}{\chi} \right)
- \frac{\pi}{2 \chi^2} 
+ \pi \Theta(-\calE_q) 
\frac{ 2\sin^2(\tfrac{\chi}{2})   - \chi \sin(\chi) }{\chi^2} \right),
\\[2.0ex]
J_2(\chi) \doteq & \; \calE_q^2 \, \left( \frac{3 \, \pi}{\chi^4} 
+ \sgn(\calE_q) \,
\left[ \frac{4 \chi - 3 \pi}{\chi^4} +
\frac{3 \, (2 - \chi^2)}{\chi^4} T(\chi) +
\frac{\chi^2 - 6}{\chi^3} \, U(\chi) \right]
- \pi \Theta(-\calE_q) \,
\frac{\partial^2}{\partial \chi^2} 
\frac{ 2\sin^2(\tfrac{\chi}{2}) - \chi \sin(\chi) }{\chi^2} \right) \,.
\end{align}
\end{subequations}
We can finally give the complete result for the 
mixing term $\Delta M$, with full account of retardation, 
as a sum over virtual states $| q\rangle$,
\begin{align}
\label{RESM}
\Delta M \doteq & 
\frac{e^2}{4 \pi} \sum_q \calE_q^4 \,
\Biggl\{ 
\langle m | \calT_1 | n \rangle \, \left[ 
\sgn(\calE_q) \,
\left( \frac{4 + \pi \chi}{2 \, \chi^3} 
- \frac{T(\chi)}{\chi^2} 
+ \frac{U(\chi)}{\chi} \right)
-\frac{\pi}{2 \chi^2} 
+ \pi \Theta(-\calE_q) \, \frac{2\sin^2(\tfrac{\chi}{2}) - \chi \, \sin(\chi)}{\chi^2} \right]
\\[2.0ex] 
& \; 
+ \langle m | \calT_2 | n \rangle 
\left[ \sgn(\calE_q) 
\left( \frac{3 \pi - 4 \, \chi}{\chi^4} 
+ \frac{3 (\chi^2 - 2)}{\chi^4} T(\chi) 
+ \frac{6 - \chi^2}{\chi^3} U(\chi) \right) 
-\frac{3 \pi}{\chi^4} + 
\pi \Theta(-\calE_q) 
\frac{\partial^2}{\partial \chi^2} 
\frac{2\sin^2(\tfrac{\chi}{2}) - \chi \sin(\chi)}{\chi^2} 
\right] \Biggr\}  \,.
\nonumber
\end{align}
The energy variable $\calE_q$ is defined with respect to the 
reference state; i.e., if one evaluates the $|m\rangle$-state
admixture to the reference state $| n \rangle$, then one sets
$\calE_q = E_q - E_n$.
For excited reference states, results for 
both $\Delta E$ given in Eq.~\eqref{RESH} 
and $\Delta M$ in Eq.~\eqref{RESM} contain 
long-range retardation tails for excited reference states,
\begin{align}
\label{ASYH}
\Delta E = & \; e^2 \,
\sum_q \Theta(-\calE_q) \,
\left[ | \langle n | \vec r_\parallel | q \rangle |^2 \,
\left( \frac{ \calE_q^2 \, \cos(2 \calE_q \calZ) }{2 \, \calZ} 
- \frac{ \calE_q \, \sin(2 \calE_q \calZ) }{4 \, \calZ^2} 
- \frac{ \cos(2 \calE_q \calZ) }{8 \, \calZ^3} \right)
\right.
\nonumber\\[2.0ex]
& \; \left. - \, | \langle n | z | q \rangle |^2 \,
\left( \frac{ \calE_q \, 
\sin(2 \, \calE_q \, \calZ) }{\calZ^2} + \frac{ \cos(2 \calE_q \calZ) }{4 \, \calZ^3} \right)
\right] - \frac{1}{8 \, \pi \, \calZ^4} \,
\left( 2 \Pi_\parallel + \Pi_\perp \right) \,,
\qquad 
\qquad 
\calZ \gg \frac{1}{\calE_q} \,,
\nonumber\\[2.0ex]
\Pi_\parallel = & \;
\frac12 \, \sum_{q, \pm} \frac{2}{\calE_q} \,
\langle n | \vec r_\parallel | q \rangle \cdot
\langle q | \vec r_\parallel | n \rangle  \,,
\qquad
\Pi_\perp =
\sum_{q, \pm} \frac{2}{\calE_q} \,
\left| \langle n | z | q \rangle \right|^2 \,,
\quad
\Pi(\omega)
= \frac{e^2}{3} \sum_{\pm} 
\left< n \left| r^i \left( \frac{1}{\calE_q \pm \omega} \right) 
r^i \right| n \right> \,,
\end{align}
where $\Pi_\parallel $ and $\Pi_\perp $ are the longitudinal 
and transverse static polarizabilities 
[for the ground state, $\Pi_\perp = \Pi_\parallel = \Pi(0)$].
The mixing term has the following long-range asymptotics,
\begin{align}
\label{ASYM}
\Delta M = & \; e^2 \sum_q \Theta(-\calE_q) \,
\langle m | {\vec r}_\parallel | q \rangle \cdot
\langle q | {\vec r}_\parallel \, z | n \rangle 
\left( -\frac{\calE_q^3 \sin(2 \calE_q \, \calZ)}{4 \calZ} 
- \frac{3\,\calE_q^2 \cos(2 \calE_q \, \calZ)}{8 \calZ^2} 
+ \frac{3 \, \calE_q \, \sin(2 \calE_q \, \calZ)}{8 \calZ^3} 
+ \frac{3\, \calE_q^4 \, \cos(2 \calE_q \, \calZ)}{16 \calZ^4} \right)
\nonumber\\[2.0ex]
& \; + e^2 \sum_q \Theta(-\calE_q) \,
\langle m | z | q \rangle 
\langle q | \vec r_\parallel^{\,2} - 2 z^2  | n \rangle 
\left( \frac{\calE_q^2 \, \cos(2 \calE_q \, \calZ)}{8 \calZ^2} 
- \frac{3 \, \calE_q \, \sin(2 \calE_q \, \calZ)}{16 \calZ^3} 
- \frac{3\, \cos(2 \calE_q \, \calZ)}{32 \calZ^4} \right)
\\[2.0ex]
& \; + \frac{e^2}{\pi \, \calZ^5} \sum_q 
\frac{1}{\calE_q}
\left( - \frac18 \,
\langle m | z | q \rangle \,
\langle q | {\vec r}_\parallel^{\,2} | n \rangle 
+ \frac14 \, \langle m | z | q \rangle \,
\langle q | z^2 | n \rangle 
+ \frac38 \, \langle m | {\vec r}_\parallel | q \rangle \cdot
\langle q | {\vec r}_\parallel \, z | n \rangle \right) 
+ \langle m | {\rm h.c.} | n \rangle  \,,
\qquad \calZ \gg \frac{1}{\calE_q} \,.
\nonumber
\end{align}
The results~\eqref{RESH} and~\eqref{RESM} will now be 
applied to metastable hydrogen.
\end{widetext}

%
%
{\em Nonretarded admixtures to metastable hydrogen.---}The results given
in Eq.~\eqref{RESH} and~\eqref{RESM} have a rather involved analytic structure.
In the short-range limit, these results can be compared to the static
interaction of the electron and proton~\cite{Ch1968,ErrorChaplik} with their
respective mirror charges.  This interaction leads to the following nonretarded
potential (from now on we set the elementary charge $e=1$),
\begin{align}
\label{VVV}
V =& \; \frac12 \, \left( - \frac{1}{2 (z + \calZ)}
+ \frac{2}{\sqrt{x^2 + y^2 + (z + 2 \calZ)^2}}
- \frac{1}{2 \calZ} \right) 
\nonumber\\[2.0ex]
=& \; - \frac{{\vec r}_\parallel^2  + 2 z^2}{16 \, \calZ^3} 
+ \frac{3 z \, ({\vec r}_\parallel^2  + 2 z^2)}{32 \, \calZ^4} + \cdots
\end{align}
where we ignore terms of order $1/\calZ^5$ and higher~\cite{factor,CTDiLa1978vol2}.
After some tedious, but straightforward algebra,
one can convince oneself that
the terms of order $\calZ^{-3}$ and $\calZ^{-4}$ are in agreement 
with the short-range asymptotics of the results given in 
Eqs.~\eqref{RESH} and~\eqref{RESM}, i.e., in the regime $\calZ \ll 1/\calE_q$, 
which is equivalent to the limit $\chi \to 0$.

For close approach of the atom to the wall, the interaction energy is 
well described by the static potential~\eqref{VVV},
which necessitates 
a diagonalization of the Schr\"{o}dinger Hamiltonian plus the 
nonretarded potential $V$ (both ``diagonal'' 
interaction and Lamb shift or fine structure terms, as well 
as ``mixing'' terms) in
the basis of the $| 2S_{1/2} \rangle$, $| 2P_{1/2} \rangle$,
and $| 2P_{3/2} \rangle$
Schr\"{o}dinger--Pauli wave functions
with magnetic projection $\mu = + 1/2$,
to form the manifestly coupled states
$| \calS_{1/2} \rangle$, $| \calP_{1/2} \rangle$, and $| \calP_{3/2} \rangle$.
We denote the (free-space) fine-structure and the Lamb shift 
interval as 
\begin{equation}
\label{LF}
\calF  = 1.66 \times 10^{-6} \, {\rm a.u.} \,,
\qquad
\calL = 1.61 \times 10^{-7} \, {\rm a.u.} \,,
\end{equation}
respectively.
According to the adiabatic theorem~\cite{BoFo1928,Ka1950,AvEl1999},
the $| \calS_{1/2} \rangle$ state eigenvector has the form
\begin{subequations}
\label{admix}
\begin{align}
| \calS_{1/2} \rangle \approx & \;
a_S \, | 2S_{1/2} \rangle + a_{\frac12} \, | 2P_{1/2} \rangle 
+ a_{\frac32} \, | 2P_{3/2} \rangle  \,,
\\[2.0ex]
a_S =& \; 1 \,,
\quad
a_{\frac12} = 
\frac{\sqrt{3}}{2} \frac{15}{\calL \, \calZ^4} \,,
\quad
a_{\frac32} = 
\sqrt{\frac{3}{2}} \frac{15}{\calF \, \calZ^4} \,,
\\[2.0ex]
1/\calL \gg & \; \calZ \gg 1/\calL^{1/4} \,,
\qquad
1/\calZ \gg \calZ \gg 1/\calF^{1/4} \,,
\end{align}
\end{subequations}
where we ignore higher-order terms in the 
expansion in inverse powers of $\calZ$.
The absolute square of the admixture is given by 
\begin{equation}
\label{Xi}
\Xi = \frac{675}{2} \, 
\left( \frac{1}{\calF^2} + \frac{1}{2 \, \calL^2} \right) \frac{1}{\calZ^8} 
= \frac{6.63 \times 10^{15}}{\calZ^8} \, {\rm a.u.} \,.
\end{equation}
The one-photon decay width of the $2P$ state 
is given as $\Gamma_{2P} = 6.27 \times 10^8 \;  \frac{{\rm rad}}{{\rm s}} 
= 1.51 \times 10^{-8} \; {\rm a.u.}$,
whereas the two-photon decay width of the $2S$ state
reads $\Gamma_{2S} = 8.229 \;  \frac{{\rm rad}}{{\rm s}} 
= 1.99 \times 10^{-16} \; {\rm a.u.}$.
The effective decay rate $\Gamma_{\rm eff}$ 
at a distance $\calZ$ is 
\begin{equation}
\label{GammaZ}
\Gamma_{\rm eff} = \Gamma_{2S} + \Gamma_{2P} \Xi 
= \left( 1.99 \times 10^{-16} 
+ \frac{1.01 \times 10^8}{\calZ^8} \right) \! {\rm a.u.}.
\end{equation}
We have $\Gamma_{\rm eff} = 2 \, \Gamma_{2S}$
for $\calZ_0 = 918 \, {\rm a.u.}$.
The leading (nonretarded) term in the 
atom-wall energy shift at this distance amounts to $-7 \calZ_0^{-3}/2 = 
-4.52 \times 10^{-9} \; {\rm a.u.}$ and approximates both the 
single-particle perturbative shift given in Eq.~\eqref{RESH} 
as well as the adiabatic energy of the 
coupled $|\calS_{1/2}\rangle$ state obtained from the 
diagonalization of the potential~\eqref{VVV}
to within 10\,\%. The atom-wall energy 
at $\calZ_0$ is equal to 
$-29.7 \, {\rm MHz}$ and thus much smaller than the Lamb shift and fine structure.

\begin{figure}[t!]
\includegraphics[width=0.91\linewidth]{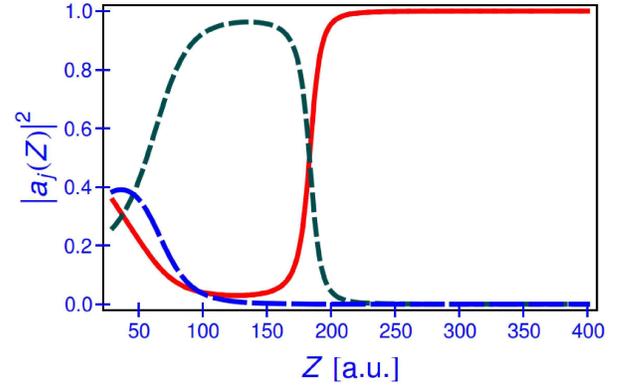}
\caption{\label{figg1} (Color.)
The modulus-squared admixtures
to the coupled $| \calS_{1/2}\rangle$ state
are obtained from a diagonalization of the
potential~\eqref{VVV} in the basis of
$| \calS_{1/2} \rangle$, $| \calP_{1/2} \rangle$, and
$| \calP_{3/2} \rangle$ states, for close approach
of the atom toward the wall. The subscript $j$
in Eq.~\eqref{admix} takes on the values
$j = S$, as well as $j = 1/2$ and $j=3/2$ and denotes
the state responsible for the admixture.
As the $| \calS_{1/2} \rangle$ state approaches the wall, the initially dominant
$| S_{1/2} \rangle$ state contribution
(solid curve, $j=S$) gradually fades
and the $| P_{1/2} \rangle$ admixture (short-dashed curve, $j=1/2$)
increases, while a significant admixture of the
$| P_{3/2} \rangle$ state (long-dashed curve, $j=3/2$),
is observed only for close approach.
The atom-wall interaction energy becomes commensurate with the
Lamb shift and fine structure at $\calZ \approx 84$ and
at $\calZ \approx 184$, respectively.}
\end{figure}

The admixture formulas for the coupled $| \calP_{1/2}\rangle$ state
reads as
\begin{subequations}
\label{admixP12}
\begin{align}
| \calP_{1/2} \rangle \approx & \;
b_S \, | 2S_{1/2} \rangle + b_{1/2} \, | 2P_{1/2} \rangle 
+ b_{3/2} \, | 2P_{3/2} \rangle  \,,
\\[2.0ex]
b_S =& \; -\sqrt{\frac{3}{4}} \frac{15}{\calL \, \calZ^4} \,,
\quad
b_{\frac32} = 
\frac{1}{2 \, \sqrt{2}} \, \frac{1}{\calF \, \calZ^3} \,,
\end{align}
\end{subequations}
and $b_{\frac12} = 1 $.
The $| \calP_{3/2} \rangle$ state reads as follows,
\begin{subequations}
\begin{align}
| \calP_{3/2} \rangle \approx & \;
c_S \, | 2S_{1/2} \rangle + c_{1/2} \, | 2P_{1/2} \rangle 
+ c_{3/2} \, | 2P_{3/2} \rangle  \,,
\\[2.0ex]
c_S =& \; -\sqrt{\frac{3}{2}} \frac{15}{\calF \, \calZ^4},
\quad
c_{\frac12} =
\frac{1}{2 \sqrt{2}} \frac{1}{(\calL + \calF) \, \calZ^3} ,
\end{align}
\end{subequations}
and of course $c_{\frac32} = 1$.
For very close approach $Z \lesssim 300$,
higher-order terms in the expansion 
of the potential $V$ [see Eq.~\eqref{VVV}]
gradually become important. 
[These are obtained by straightforward expansion
of the potential~\eqref{VVV}.]
Numerically determined admixtures of the
coupled $|\calS_{1/2}\rangle$ are
given in Fig.~\ref{figg1},
they do not follow the asymptotic
formulas for very close approach.

%
%
{\em Long-range tails.---}The oscillatory 
repulsive-attractive dominant
term in the long-range limit of the energy shift,
for the $2S$ level, goes as [see Eqs.~\eqref{RESH} and~\eqref{ASYH}],
\begin{equation}
\Delta E_{2S} \sim 
\frac{ 9 \calL^2 \, \cos(2 \calL \, \calZ) }{2 \calZ} \,,
\qquad \calZ \gg \frac{1}{\calL}\,,
\end{equation}
where we have isolated the leading term from 
Eq.~\eqref{ASYH}, setting $\calE_q = -\calL$ and 
carrying the summation over the virtual levels
$|q \rangle = | 2 P_{1/2} \rangle$ with 
magnetic projections $\mu = \pm 1/2$.
Somewhat surprisingly, the oscillatory 
terms in Eq.~\eqref{ASYM} vanish for virtual
$| 2 P_{1/2} \rangle$ states, so that the long-range
coupling to the lower-lying $P$ state vanishes.
The leading terms in the 
long-range asymptotics of the admixture 
coefficients read as follows [see Eq.~\eqref{admix}],
\begin{subequations}
\label{resx}
\begin{align}
\label{resxa}
a_{1/2} \sim & \; \frac{3 \, \sqrt{3}}{\pi \, \calL \, \calF \, \calZ^5} \,, 
\qquad \qquad \calZ \gg \frac{1}{\calL} \,,
\\[2ex]
\label{resxb}
a_{3/2} \sim & \; - \sqrt{\frac{3}{2}} \, 
\frac{3 \, \calL^3}{\calF \, \calZ} \,
\sin(2 \calL \, \calZ) \,,
\qquad \calZ \gg \frac{1}{\calL} \,.
\end{align}
\end{subequations}
The long-range asymptotic tail of the $P_{3/2}$-state 
admixture has an oscillatory ($1/\calZ$)-form
[see Eqs.~\eqref{ASYM} and~\eqref{resxb}].
If this tail were not suppressed by the prefactor $\calL^3/\calF$,
then it could have easily provided a theoretical
explanation for the Sokolov 
effect~\cite{SoYaPa1994,KaKaKuPoSo1996,SoYaPaPc2002,SoEtAl2005},
because the $(1/\calZ)$-interaction has the required 
functional form to describe a super-long-range term.
The tail is created by virtual 
$| q \rangle = | 2P_{1/2} \rangle$ states in Eq.~\eqref{ASYM},
which are energetically lower than the reference $|2S\rangle$ state.
The prefactor of the super-long-range 
tail of the admixture term depends on details of the 
spectrum of the atomic system and could be larger for 
other atoms. For the $P_{1/2}$-state admixture (term $a_{1/2}$),
retardation changes the $1/\calZ^4$ asymptotics
for short range to a $1/\calZ^5$ asymptotics
at long range.  A full QED treatment of the admixture terms
is required for both results recorded in 
Eqs.~\eqref{resxa} and~\eqref{resxb}.

%
%
{\em Conclusions.---}We can safely conclude that the 
curious observations reported 
in~\cite{SoYaPa1994,KaKaKuPoSo1996,SoYaPaPc2002,SoEtAl2005}
regarding super-long-range $2S$--$2P$ mixing terms 
near metal surfaces cannot find an explanation
in terms of a long-range effect involving quantum fluctuations.
Both the energy shift~\eqref{ASYH} as well as the 
mixing term~\eqref{ASYM} have long-range tails
proportional to $1/\calZ$, but the energy numerator for the 
$2S$--$2P_{1/2}$ transition is so small (Lamb shift, a
$30 \,{\rm cm}$ wavelength transition) 
that the region in which the $1/\calZ$ terms dominate
is restricted to excessively large atom-wall separations
where the single power of $\calZ$ in the denominator
is sufficient to make the interaction energy and admixture terms 
negligible. (We should add that 
the inclusion of additional mirror charges 
in a cavity as opposed to a wall can be 
taken into account, in the short-range limit,
by summing the mirror charge interactions into a 
generalized Riemann zeta function~\cite{LuRa1985}
and therefore cannot change the order-of-magnitude of the 
admixture terms.)

If the observations reported 
in Refs.~\cite{SoYaPa1994,KaKaKuPoSo1996,SoYaPaPc2002,SoEtAl2005} had found a 
natural explanation in terms of a QED effect, then 
this might have had significant implications for 
a typical atomic beam apparatus~\cite{FiEtAl2004}
used in high-precision spectroscopy of atoms,
potentially shifting the frequency of transitions involving $2S$ 
atoms in a narrow tube.
For atom-wall separations smaller than $1000$~Bohr radii,
substantial admixture terms are found,
and the $1/\calZ^8$ scaling of the 
effective $2S$ decay rate predicted by Eq.~\eqref{Xi} could be 
tested against an experiment.
The clarification of the parity-breaking admixture terms
also is important for other precision measurements in 
atomic physics which involve metastable states, 
such as EDM and weak-interaction experiments~\cite{WuMuJe2012prx,BeGaNa2007paper1,%
BeGaNa2007paper2,BuDMCoZo1994,NgEtAl1997,WeLeBu2013}.
The fully retarded expression for the mixing term,
given in Eq.~\eqref{ASYM}, 
formulates higher-order QED corrections to 
atom-wall interactions beyond dipole order.
Generalization of the formulas to, e.g.,
the $2^3 S_1$ mestable state of helium is straightforward.
One just sums the interactions over the electron coordinates.

\begin{acknowledgments}

Support from the National
Science Foundation (grants PHY--1068547 and PHY--1403973) 
and helpful conversations with M. M. Bush are  gratefully acknowledged.

\end{acknowledgments}


\begin{thebibliography}{10}

\bibitem{LJ1932}
J.~E. Lennard-Jones, Trans. Faraday Soc. {\bf 28},  334  (1932).

\bibitem{Ba1940}
J. Bardeen, Phys. Rev. {\bf 58},  727  (1940).

\bibitem{CaPo1948}
H.~B.~G. Casimir and D. Polder, Phys. Rev. {\bf 73},  360  (1948).

\bibitem{Li1955}
E.~M. Lifshitz, Zh. \'{E}ksp. Teor. Fiz. {\bf 29},  94  (1955),
  [Sov.~Phys.~JETP {\bf 2}, 73 (1956)].

\bibitem{WySi1984}
J.~M. Wylie and J.~E. Sipe, Phys. Rev. A {\bf 30},  1185  (1984).

\bibitem{WySi1985}
J.~M. Wylie and J.~E. Sipe, Phys. Rev. A {\bf 32},  2030  (1985).

\bibitem{BaDo1987}
A.~O. Barut and J.~P. Dowling, Phys. Rev. A {\bf 36},  2550  (1987).

\bibitem{BaDo1990}
A.~O. Barut and J.~P. Dowling, Phys. Rev. A {\bf 41},  2284  (1990).

\bibitem{Mi1994}
P.~W. Milonni, {\em \relax{The Quantum Vacuum}} (Academic Press, San Diego,
  1994).

\bibitem{LaDKJe2010pra}
G. \L{}ach, M. DeKieviet, and U.~D. Jentschura, Phys. Rev. A {\bf 81},  052507
  (2010).

\bibitem{BoEtAl2001}
M. Boustimi, B. Viaris~de Lesegno, J. Baudon, J. Robert, and M. Ducloy, Phys.
  Rev. Lett. {\bf 86},  2766  (2001).

\bibitem{KaEtAl2006epl}
J.-C. Karam {\it et~al.}, Europhys. Lett. {\bf 74},  36  (2006).

\bibitem{SoYaPa1994}
Y.~L. Sokolov, V.~P. Yakovlev, and V.~G. Pal'chikov, Phys. Scr. T {\bf 49},  86
   (1994).

\bibitem{KaKaKuPoSo1996}
B.~B. Kadomtsev, M.~B. Kadomtsev, Y.~A. Kucheryaev, Y.~L. Podogov, and Y.~L.
  Sokolov, Phys. Scr. T {\bf 54},  156  (1996).

\bibitem{SoYaPaPc2002}
Y.~L. Sokolov, V.~P. Yakovlev, V.~G. Pal'chikov, and Y.~A. Pchelin, Eur. Phys.
  J. D {\bf 20},  27  (2002).

\bibitem{SoEtAl2005}
Y.~L. Sokolov, V.~P. Yakovlev, V.~G. Pal'chikov, and Y.~A. Pchelin, Pis'ma v
  ZhETF {\bf 81},  780  (2005), [JETP Lett. {\bf 81}, 644 (2005)].

\bibitem{Ba1970}
G. Barton, Proc. Roy. Soc. London, Ser. A {\bf 320},  251  (1970).

\bibitem{BaBa1972a}
M. Babiker and G. Barton, Proc. Roy. Soc. London, Ser. A {\bf 326},  255
  (1978).

\bibitem{BaBa1972b}
M. Babiker and G. Barton, Proc. Roy. Soc. London, Ser. A {\bf 326},  277
  (1972).

\bibitem{Ba1974}
G. Barton, J. Phys. B {\bf 7},  2134  (1974).

\bibitem{Ba1979}
G. Barton, Proc. Roy. Soc. London, Ser. A {\bf 367},  117  (1979).

\bibitem{Fo1979}
L.~H. Ford, Proc. Roy. Soc. London, Ser. A {\bf 368},  311  (1979).

\bibitem{Ba1987a}
G. Barton, Proc. Roy. Soc. London, Ser. A {\bf 410},  141  (1987).

\bibitem{Ba1987b}
G. Barton, Proc. Roy. Soc. London, Ser. A {\bf 410},  175  (1987).

\bibitem{Ba1988}
G. Barton, Phys. Scr. T {\bf 21},  11  (1988).

\bibitem{Ba1997}
G. Barton, Proc. Roy. Soc. London, Ser. A {\bf 453},  2461  (1997).

\bibitem{PoZi1959}
E.~A. Power and S. Zienau, Phil. Trans. R. Soc. Lond. A {\bf 251},  427
  (1959).

\bibitem{Pa2004}
K. Pachucki, Phys. Rev. A {\bf 69},  052502  (2004).

\bibitem{Ko1978prl}
D.~H. Kobe, Phys. Rev. Lett. {\bf 40},  538  (1978).

\bibitem{JeKe2004aop}
U.~D. Jentschura and C.~H. Keitel, Ann. Phys. (N.Y.) {\bf 310},  1  (2004).

\bibitem{Le1981qed}
T.~D. Lee, {\em \relax{Particle Physics and Introduction to Field Theory}}
  (Harwood Publishers, Newark, NJ, 1981).

\bibitem{Sa1967Adv}
J.~J. Sakurai, {\em \relax{Advanced Quantum Mechanics}} (Addison-Wesley,
  Reading, MA, 1967).

\bibitem{Ch1968}
A.~V. Chaplik, Zh. \'{E}ksp. Teor. Fiz. {\bf 54},  332  (1968), [JETP {\bf 27},
  178 (1968)].

\bibitem{ErrorChaplik}
The nonretarded potential used in Ref.~\cite{Ch1968} for the analysis of a
  related problem, is incomplete as it neglects the interaction of the proton
  with its mirror charge.

\bibitem{factor}
The well-known prefactor $1/2$ stands because the electric field is zero inside
  the conductor (see Chap.~11 of Ref.~\cite{CTDiLa1978vol2}).

\bibitem{CTDiLa1978vol2}
C. Cohen-Tannoudji, B. Diu, and F. Lalo$\ddot{e}$, {\em \relax{Quantum
  Mechanics (Volume 2)}}, 1 ed. (J. Wiley \& Sons, New York, 1978).

\bibitem{BoFo1928}
M. Born and V.~A. Fock, Z. Phys. A {\bf 51},  165  (1928).

\bibitem{Ka1950}
T. Kato, J. Phys. Soc. Jpn. {\bf 5},  435  (1950).

\bibitem{AvEl1999}
J.~E. Avron and A. Elgart, Commun. Math. Phys. {\bf 203},  445  (1999).

\bibitem{LuRa1985}
C.~A. L\"{u}tken and F. Ravndal, Phys. Rev. A {\bf 31},  2082  (1985).

\bibitem{FiEtAl2004}
M. Fischer {\it et~al.}, Phys. Rev. Lett. {\bf 92},  230802  (2004).

\bibitem{WuMuJe2012prx}
B.~J. Wundt, C.~T. Munger, and U.~D. Jentschura, Phys. Rev. X {\bf 2},  041009
  (2012).

\bibitem{BeGaNa2007paper1}
T. Bergmann, T. Gasenzer, and O. Nachtmann, Eur. Phys. J. D {\bf 45},  197
  (2007).

\bibitem{BeGaNa2007paper2}
T. Bergmann, T. Gasenzer, and O. Nachtmann, Eur. Phys. J. D {\bf 45},  211
  (2007).

\bibitem{BuDMCoZo1994}
D. Budker, D. DeMille, E.~D. Commins, and M.~S. Zolotorev, Phys. Rev. A {\bf
  50},  132  (1994).

\bibitem{NgEtAl1997}
A.~T. Nguyen, D. Budker, D. DeMille, and M. Zolotorev, Phys. Rev. A {\bf 56},
  3453  (1997).

\bibitem{WeLeBu2013}
C.~T.~M. Weber, N. Leefer, and D. Budker, Phys. Rev. A {\bf 88},  062503
  (2013).

\end{thebibliography}
\end{document}